\documentclass[letter]{aa} 

\usepackage{graphicx}
\usepackage{placeins}
\usepackage{float}
\usepackage{txfonts}
\usepackage{xcolor}
\newcommand{\chg}[1]{{\color{black}{#1}}}

\begin{document}

   \title{Missing metals in DQ stars: A simple explanation}

   \author{Simon Blouin
          \inst{1}
          }

   \institute{Department of Physics and Astronomy, University of Victoria, Victoria, BC V8W 2Y2, Canada\\
              \email{sblouin@uvic.ca}
             }

   \date{\today}
 
  \abstract{
  Classical DQ stars are white dwarfs whose atmospheres contain detectable traces of carbon brought up to the surface by a convective dredge-up process. Intriguingly, unlike other white dwarf spectral classes, DQ stars virtually never show signs of external pollution by elements heavier than carbon. In this Letter, we solve this long-standing problem by showing that the absence of detectable external pollution in DQ stars is naturally explained by the impact of metal accretion on the atmospheric structure of the white dwarf. A DQ star that accretes heavy elements sees its atmospheric density decrease, which leads to a sharp drop in the molecular carbon abundance and a strong suppression of the C$_2$ Swan bands. We show that a typical DQ star that accretes heavy elements from planetary material generally transforms directly into a DZ star.
  }

   \keywords{planetary systems -- stars: abundances -- stars: atmospheres -- white dwarfs
               }

   \maketitle
%

\section{Introduction}
Cool ($T_{\rm eff} \lesssim 10{,}000\,{\rm K}$) helium-atmosphere white dwarfs often have detectable traces of carbon in their atmospheres. Such carbon-polluted white dwarfs are known as DQ stars. Spectroscopically, DQ white dwarfs are recognized by the presence of C$_2$ molecular bands from the Swan transitions in the optical region. Given the relentless action of gravitational settling, heavy elements such as carbon are a priori not expected to be present in white dwarf atmospheres. But many physical processes can compete with gravitational settling. In the case of cool DQ stars, the presence of carbon is understood as the result of convective dredge-up from the deep interior \citep{pelletier1986,macdonald1998,althaus2005,dufour2005,koester2020}. Under reasonable modelling assumptions, the most recent evolutionary calculations agree remarkably well with the observed carbon abundance pattern of DQ stars \citep{bedard2022}.

The atmospheres of cool helium-rich white dwarfs are also frequently polluted by heavier elements, such as Ca, Mg and Fe. Those are known as DZ stars. The origin of heavy elements in DZ white dwarfs is now convincingly explained by the recent or ongoing accretion of rocky debris \citep[e.g.,][]{jura2014,farihi2016}. Many independent observations all converge to this explanation, including the abundance patterns observed in DZ atmospheres \citep{zuckerman2007,klein2010,gansicke2012,doyle2019,harrison2021}, the identification of circumstellar disks \citep{rocchetto2015,wilson2019,manser2020}, the discovery of planetary debris by the transit method \citep{vanderburg2015,vanderbosch2020,vanderbosch2021} and the detection of X-rays from the accretion process itself \citep{cunningham2022}.

Approximately 15\% of all helium-atmosphere white dwarfs are DZs \citep{mccleery2020,hollands2022}. Under the assumption that the DQ phenomenon is independent from the external accretion of rocky planetesimals, about 15\% of DQ white dwarfs should also have atmospheres polluted by elements heavier than carbon. But it is well established that DQZ/DZQ stars (white dwarfs showing both carbon features and absorption lines from heavier elements) are exceedingly rare. Only about 2\% of DQs show the presence of heavier elements in their spectra \citep{coutu2019,farihi2022}. To add to this mystery, the inferred external accretion rates of the few known DQZs are orders of magnitude smaller than those typical of DZ stars. To solve this long-standing conundrum, \cite{farihi2022} recently suggested that DQ stars are the product of a binary evolution that has altered their circumstellar environments in a way that prevents the pollution of the white dwarf.

In this Letter, we show that this hypothesis is not necessary. We demonstrate that the C$_2$ Swan bands of typical DQ white dwarfs are efficiently suppressed if the atmosphere is polluted by an amount of metals typical of that observed in DZ stars. This naturally explains both the paucity of DQZ white dwarfs and the systematically low accretion rates inferred for known DQZ stars. In Section~\ref{sec:suppress}, we present DQZ model atmosphere calculations that show how the Swan bands can disappear if the atmosphere is moderately polluted by heavy elements following the accretion of rocky debris. We then detail in Section~\ref{sec:physics} the physical mechanism that leads to this Swan bands suppression. In Section~\ref{sec:implications}, we compare our DQZ models to observations to demonstrate that the rarity of DQZ stars can naturally be explained by this process. Finally, our conclusions are stated in Section~\ref{sec:conclusion}.

\section{Metals can suppress the C$_2$ Swan bands}
\label{sec:suppress}
The possibility of Swan bands suppression in DQZ stars was recently studied by \cite{hollands2022}. Their Figure~3 shows how their models predict that the Swan bands disappear if a sufficient quantity of polluting metals is added to the atmosphere while keeping the other parameters constant \chg{(specifically, $T_{\rm eff}=8100\,{\rm K}$, $\log g=8.02$, $\log\,{\rm H/He}=-3.2$\footnote{In this work, $\log\,{\rm X/He}$ corresponds to the logarithm of the number abundance ratio, $\log_{10}n_{\rm \scriptscriptstyle X}/n_{\rm \scriptscriptstyle He}$.} and $\log\,{\rm C/He}=-4.5$)}. However, \cite{hollands2022} argue that this disappearance of the Swan bands cannot explain the extreme rarity of DQZ stars, since they find that the Swan bands are suppressed only at extreme levels of metal pollution. Therefore, they conclude that while the Swan bands can be sufficiently inhibited to become undetectable in some strongly polluted white dwarfs, they are generally too weakly suppressed to explain the dearth of DQZ white dwarfs.

\begin{figure}
	\includegraphics[width=\columnwidth]{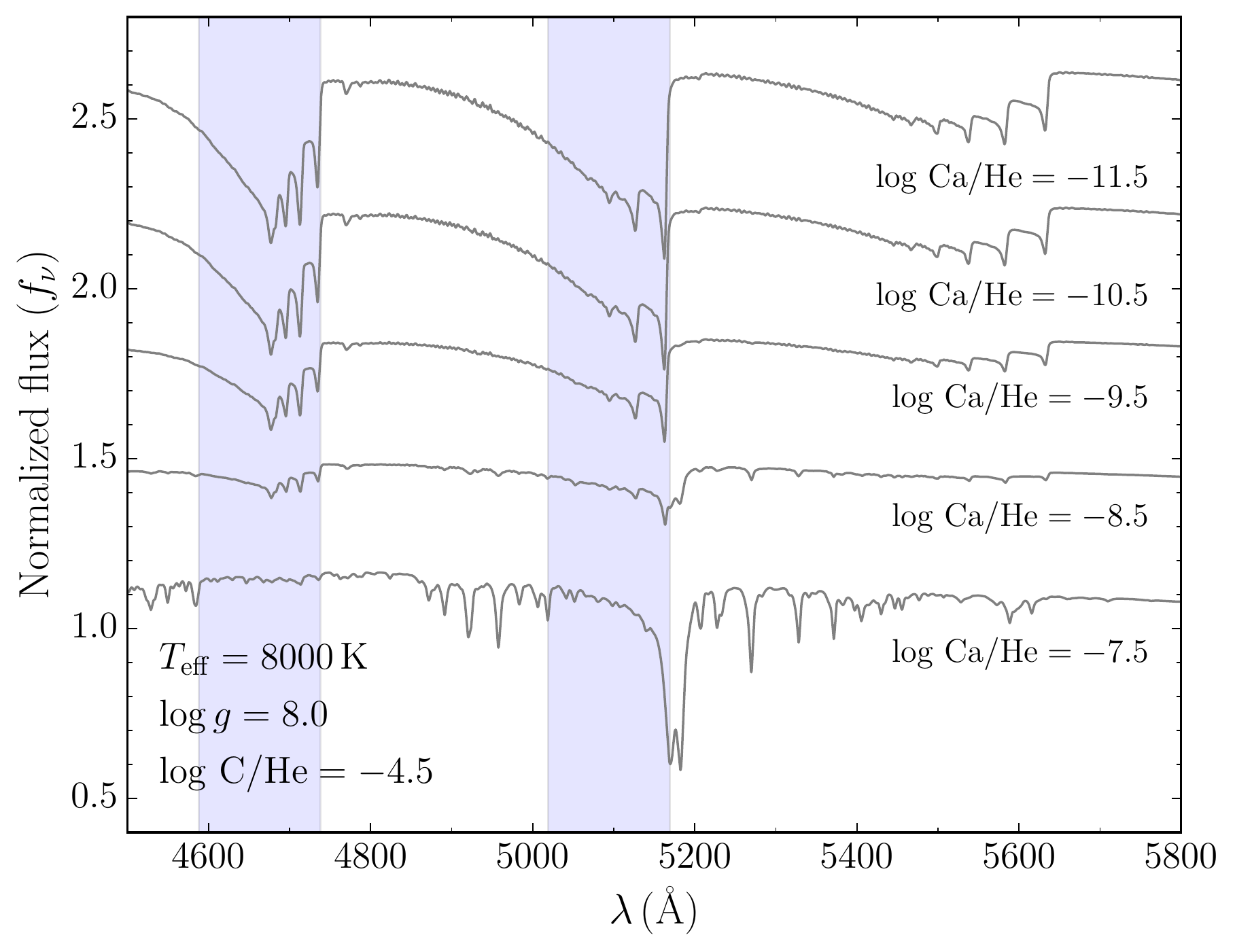}
    \caption{Synthetic spectra for a white dwarf with $T_{\rm eff}=8000\,$K, $\log g=8$, $\log\,{\rm C/He}=-4.5$ and varying amounts of polluting metals. No hydrogen trace was included. The blue regions indicate the location of the two strongest C$_2$ Swan bands.}
    \label{fig:demo_C45}
\end{figure}

\chg{In Figure~\ref{fig:demo_C45} we repeat the exercise of \cite{hollands2022} using our own state-of-the-art model atmospheres \citep{blouin2018,blouin2018b,blouin2019c}. We examine how the strength of the Swan bands changes for a normal-mass white dwarf with  $T_{\rm eff} = 8000\,{\rm K}$ and $\log\,{\rm C/He}=-4.5$ as a function of the external pollution level $\log\,{\rm Ca/He}$\footnote{We only quote the Ca abundance, but all metals from N to Cu are included with chondritic abundance ratios with respect to Ca. This applies to all models presented in this work.} (no hydrogen is included). These atmospheric parameters are similar to those used in Figure~3 of \cite{hollands2022} but there are small differences since in that work the hydrogen and individual metal abundances were adjusted to the specific case of SDSS~J095645.12+591240.7. We qualitatively replicate their finding that a high metal pollution of $\log\,{\rm Ca/He} \gtrsim -8.5$ is required to strongly suppress the Swan bands at that temperature and carbon abundance.} This good agreement between independent calculations indicates that differences with respect to the constitutive physics of our atmosphere code and that used in \citealt{hollands2022} \citep{koester2010} are negligible in the context of this work.

\begin{figure}
	\includegraphics[width=\columnwidth]{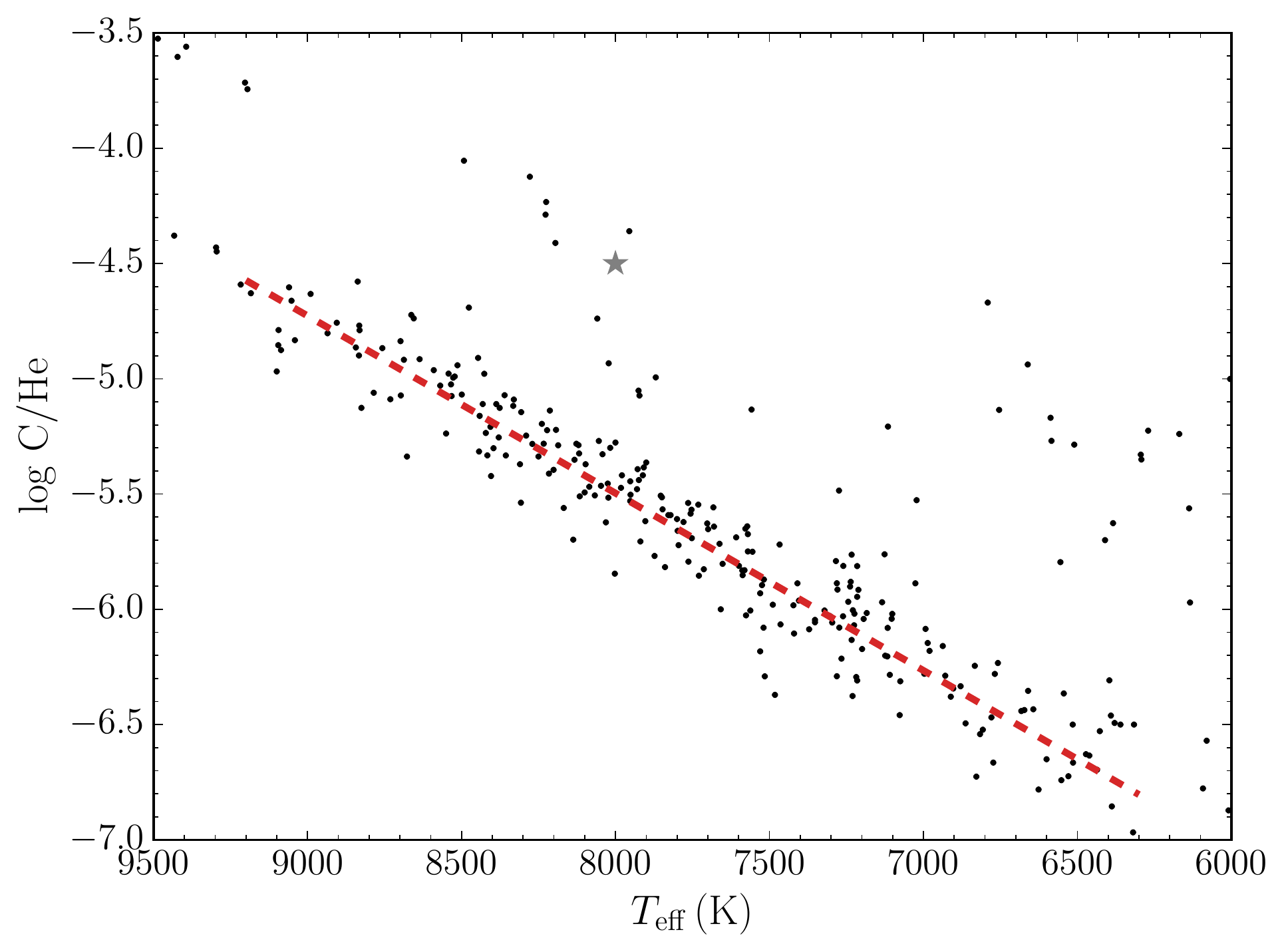}
    \caption{Photospheric carbon abundance of DQ white dwarfs as a function of effective temperature (data taken from \protect\citealt{coutu2019} and \protect\citealt{blouin2019d}). The vast majority of DQ white dwarfs follow a clear sequence in this plane as indicated by the dashed red line.}
    \label{fig:DQscatter}
\end{figure}

For their analysis, \cite{hollands2022} chose a fixed carbon abundance of $\log\,{\rm C/He}=-4.5$. This is not the ideal value to study the Swan bands suppression: DQ white dwarfs at 8000\,K typically have 10 times less carbon than that. This is shown in Figure~\ref{fig:DQscatter}, where the carbon abundances of a large sample of DQ white dwarfs are plotted as a function of $T_{\rm eff}$. At 8000\,K, $\log\,{\rm C/He}=-4.5$ \chg{(identified as a grey star)} lies a full one dex above the sequence on which the vast majority of DQs are found. While some stars are in this upper region, apparently forming a continuous sequence with the ``warm'' DQs found above $10{,}000\,$K \citep{coutu2019,koester2019}, they likely have a different origin than the classical DQs, possibly being the descendants of the Hot DQ white dwarfs. In short, the analysis presented in Figure~\ref{fig:demo_C45} of this work and in Figure~3 of \cite{hollands2022} is not representative of the typical DQ white dwarf. 

\begin{figure}
	\includegraphics[width=\columnwidth]{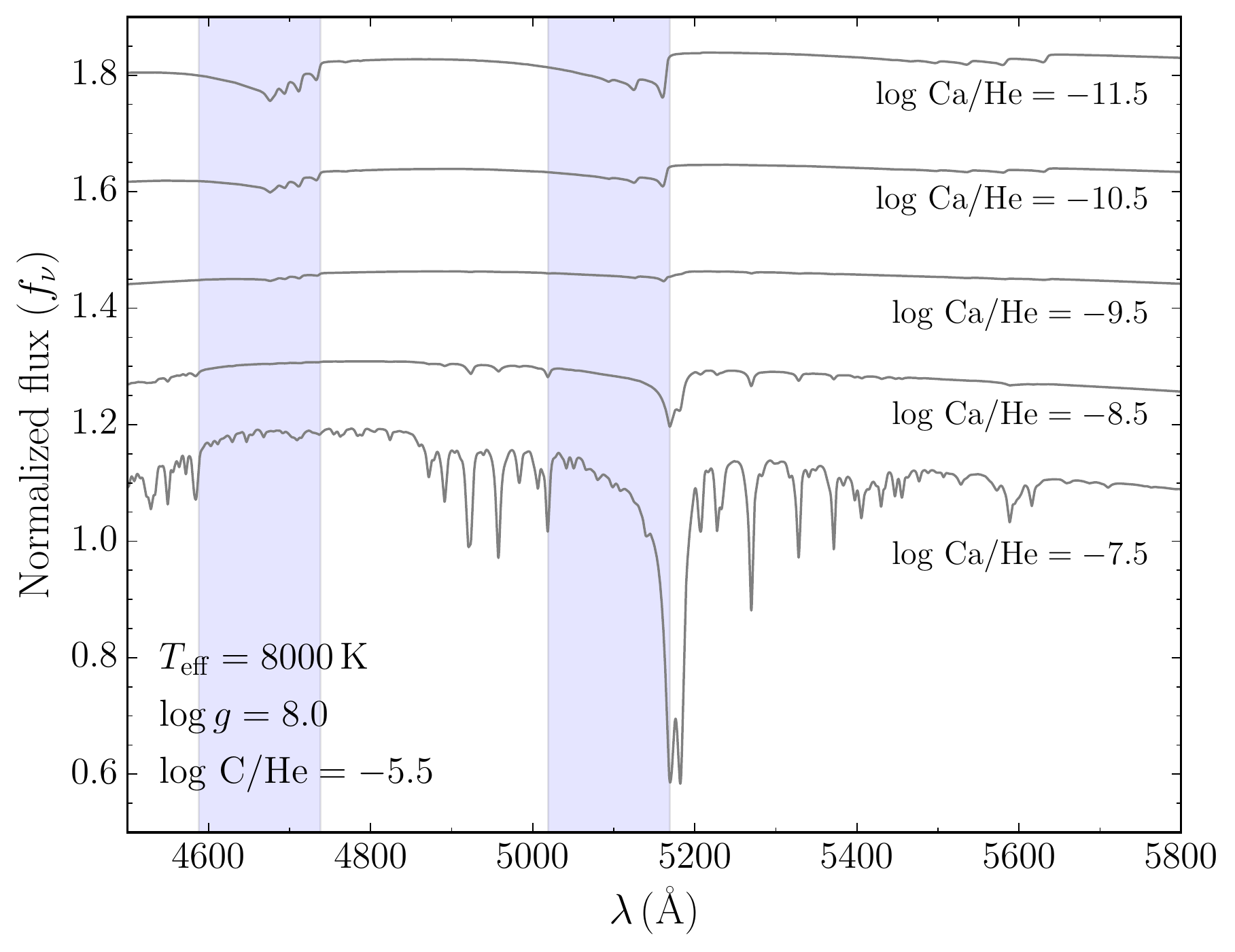}
    \caption{Same as Figure~\ref{fig:demo_C45} but this time assuming $\log\,{\rm C/He}=-5.5$. Note the different vertical scale compared to Figure~\ref{fig:demo_C45}.}
    \label{fig:demo_C55}
\end{figure}

We now repeat the exercise of \cite{hollands2022}, but this time using a carbon abundance of $\log\,{\rm C/He}=-5.5$, a representative composition for a 8000\,K DQ. Figure~\ref{fig:demo_C55} reveals that the weaker Swan bands are more easily erased from the spectrum when metals are added. The C$_2$ bands are very shallow even at the relatively low pollution level of $\log\,{\rm Ca/He}=-10.5$. This result is hardly surprising (if there is less carbon, then surely the Swan bands are weaker for any given metal pollution level), but it reopens the door to the idea that the dearth of DQZ white dwarfs could be explained by this simple effect. To test this idea, we need to compare our synthetic spectra to observations. This is the subject of Section~\ref{sec:implications}; for now, we turn to the identification of the physical mechanism responsible for the Swan bands suppression.

\section{Physical explanation}
\label{sec:physics}
To identify the reason behind the disappearance of the C$_2$ bands following the addition of metals, we look in more details at the $\log\,{\rm Ca/He}=-11.5$ and $\log\,{\rm Ca/He}=-9.5$ models of Figure~\ref{fig:demo_C55}. The first thing to establish is at which Rosseland optical depths ($\tau_R$) the Swan bands are formed in the vertical stratification of the atmosphere. To answer this question, we have computed a series of model spectra where the Swan bands opacity was artificially ignored at certain optical depths (Figure~\ref{fig:taucontrib}). From this exercise, we conclude that at this effective temperature and carbon abundance, the Swan bands are mostly formed in the $0.001 \lesssim \tau_R \lesssim 0.03$ range.

\begin{figure}
	\includegraphics[width=\columnwidth]{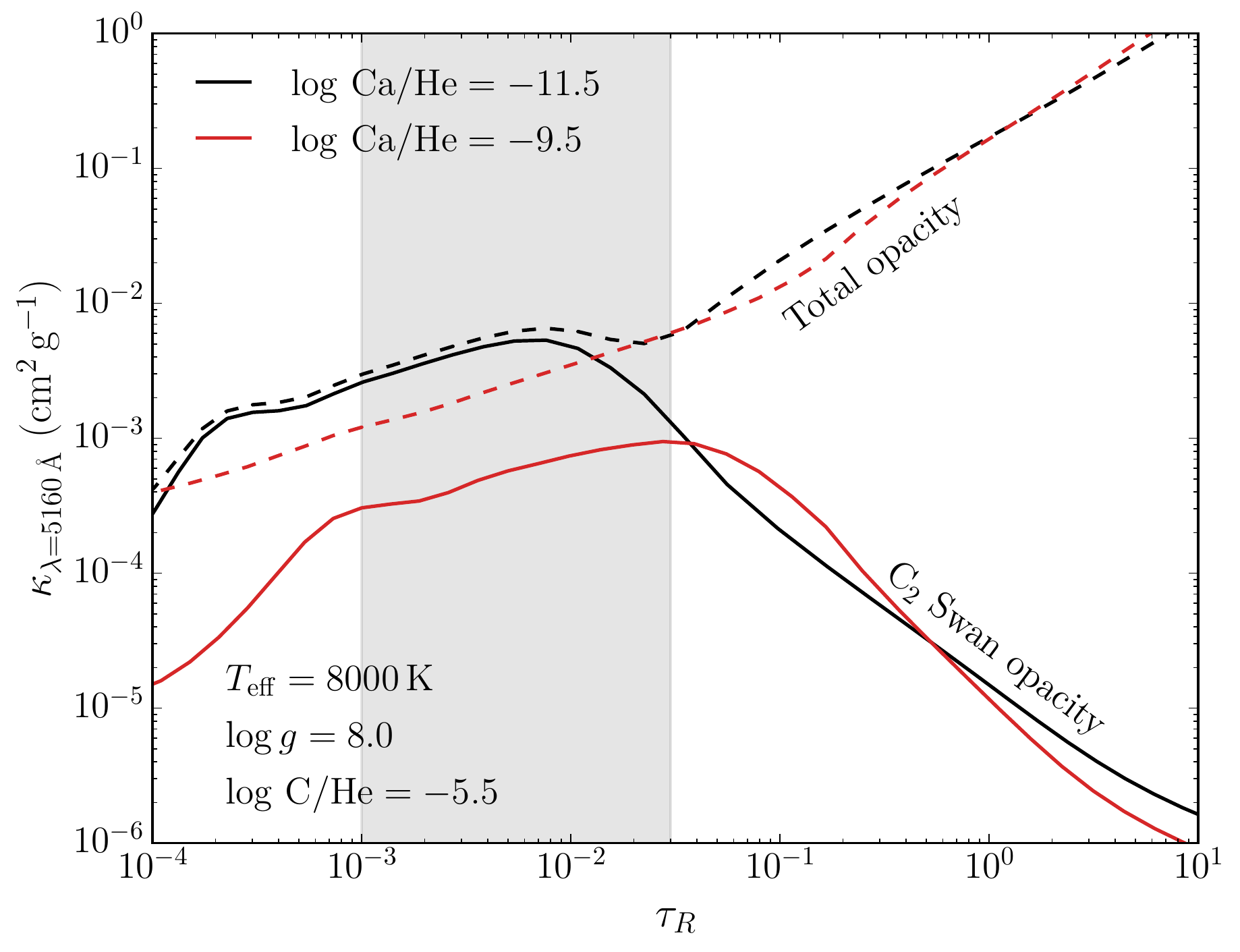}
    \caption{Opacity at $5160\,\textrm{\AA}$ as a function of Rosseland optical depth for DQZ white dwarfs with $T_{\rm eff}=8000\,$K, $\log g=8$, $\log\,{\rm C/He}=-5.5$ and different Ca/He values (black lines correspond to $\log\,{\rm Ca/He}=-11.5$ and red lines to $\log\,{\rm Ca/He}=-9.5$). The dashed lines show the total opacity while the solid lines show the opacity due to the Swan bands only. $5160\,\textrm{\AA}$ corresponds to the wavelength where the Swan bands are strongest. The grey region indicates where the Swan bands are formed.}
    \label{fig:kappa}
\end{figure}

Equipped with this information, we can now compare the $\log\,{\rm Ca/He}=-11.5$ and $\log\,{\rm Ca/He}=-9.5$ models to see how they differ. Figure~\ref{fig:kappa} compares their opacities as a function of $\tau_R$ at $\lambda=5160\,\textrm{\AA}$ (which corresponds to the wavelength where the Swan bands are strongest). It shows that the total opacity at a given $\tau_R$ (dashed lines) does not change much following the hundredfold increase in external metal pollution. In contrast, the C$_2$ Swan opacity (solid lines) decreases by a factor $\simeq 10$ in the region where the Swan bands are formed ($0.001 \lesssim \tau_R \lesssim 0.03$). Therefore, the suppression of the Swan bands seen in Figure~\ref{fig:demo_C55} is not caused by a masking effect from other opacity sources but rather by a decrease of the C$_2$ opacity itself. 

By inspecting our model structures, we found that this decline of the Swan opacity is directly caused by a tenfold decrease of the C$_2$ number density in the same region of the atmosphere (Figure~\ref{fig:C2abn}). This decrease is in turn largely explained by a $\simeq 2.5$x decrease of the total density between the $\log\,{\rm Ca/He}=-11.5$  and $\log\,{\rm Ca/He}=-9.5$ models in the relevant atmospheric layers ($0.001 \lesssim \tau_R \lesssim 0.03$, Figure~\ref{fig:structure}). The reason for this decline is well documented \citep{provencal2002,dufour2005,bergeron2019}. Metals act as electron donors that increase the total opacity of helium-dominated atmospheres mostly through an increase of the He$^-$ free-free opacity, which dominates the total atmospheric opacity in those stars \citep[][Figure~18]{saumon2022}. Due to this opacity increase, a given optical depth $\tau_R$ is attained higher up in the atmosphere, at a lower pressure.

To understand how this density decline affects the molecular carbon density, consider the C$_2$ dissociation equilibrium equation. Neglecting nonideal effects, it is given by
\begin{equation}
    \frac{n_{\rm \scriptscriptstyle C}^2}{n_{{\rm \scriptscriptstyle C}_2}} = \frac{Q_{\rm \scriptscriptstyle C}^2 (T)}{Q_{{\rm \scriptscriptstyle  C_2}} (T)} \left( \frac{ 2\pi m_{\rm \scriptscriptstyle C}^2 k_B T}{m_{{\rm \scriptscriptstyle C_2}} h^2}\right)^{3/2} e^{-D_0 / k_B T} \equiv f(T),
    \label{eq:C2}
\end{equation}
where $n_i$ is a number density, $Q_i(T)$ is a partition function, $m_i$ is a mass, $k_B$ is the Boltzmann constant, $h$ is the Planck constant and $D_0=6.21\,{\rm eV}$ is the dissociation energy of the C$_2$ molecule. At a fixed temperature and given C/He, we have
\begin{equation}
    n_{{\rm \scriptscriptstyle C}_2} \propto n_{\rm \scriptscriptstyle C}^2 \propto  \rho^2,
\end{equation}
meaning that the 2.5x decrease in mass density noted above translates into a $\simeq 6$x reduction of the C$_2$ density. 

In addition to this change in the density structure, the temperature profile is also affected (Figure~\ref{fig:structure}). This is a second-order but still important effect to explain the shift in molecular carbon abundances. In the region of interest ($0.001 \lesssim \tau_R \lesssim 0.03$), the temperature rises by $\simeq 400\,$K following the increase in metal pollution. This in turn translates into a twofold increase of the right-hand side of Equation~\eqref{eq:C2} (which we have plotted in Figure~\ref{fig:fT} for reference). Combining this effect with the density effect described in the previous paragraph we see how moderate changes to the density and temperature stratifications explain the tenfold decrease of the C$_2$ density in the Swan bands forming region of the atmosphere. We can conclude that the suppression of the Swan bands described in Section~\ref{sec:suppress} is due to changes to the atmospheric structure that tilt the scale in favour of dissociation in the C$_2$/C equilibrium equation.

\section{Implications for DQZ white dwarfs}
\label{sec:implications}

In the previous sections, we have seen how and why the accretion of metals on DQ white dwarfs can suppress their C$_2$ Swan bands. This suggests that DQ white dwarfs may transform directly into DZs (and not DQZs) following the accretion of rocky debris, which would naturally explain the apparent dearth of DQZ stars. To test this hypothesis, we turn to the Sloan Digital Sky Survey (SDSS) DZ sample of \cite{dufour2007}. This sample contains 72 DZ white dwarfs cooler than 9000\,K (a temperature range where the classical DQ sequence is well populated, see Figure~\ref{fig:DQscatter}) that have been recently reanalysed by \cite{coutu2019} using Gaia parallaxes and updated model atmospheres. Under the assumption that DQ white dwarfs descend from stars that hosted planetary systems similar to those hosted by the progenitors of non-DQ white dwarfs, we should expect that \chg{$11 \pm 3$} stars in this sample were DQs before accreting planetary material \citep[given that $\simeq 15$\% of cool helium-atmosphere white dwarfs are DQs,][]{mccleery2020}. \chg{The probability that none was a DQ is a priori $\lesssim 10^{-5}$.} The fact that Swan bands are detected in none of the 72~objects can mean two things: either the progenitors of DQs actually had different planetary system architectures after all \citep{farihi2022} or DQs generally transform directly into DZs after the accretion of metals. Of course, those two scenarios are not mutually exclusive, but if DQs do transform directly into DZs, then the need for the first scenario lessens considerably.

\begin{figure*}[ht]
    \centering
	\includegraphics[width=1.85\columnwidth]{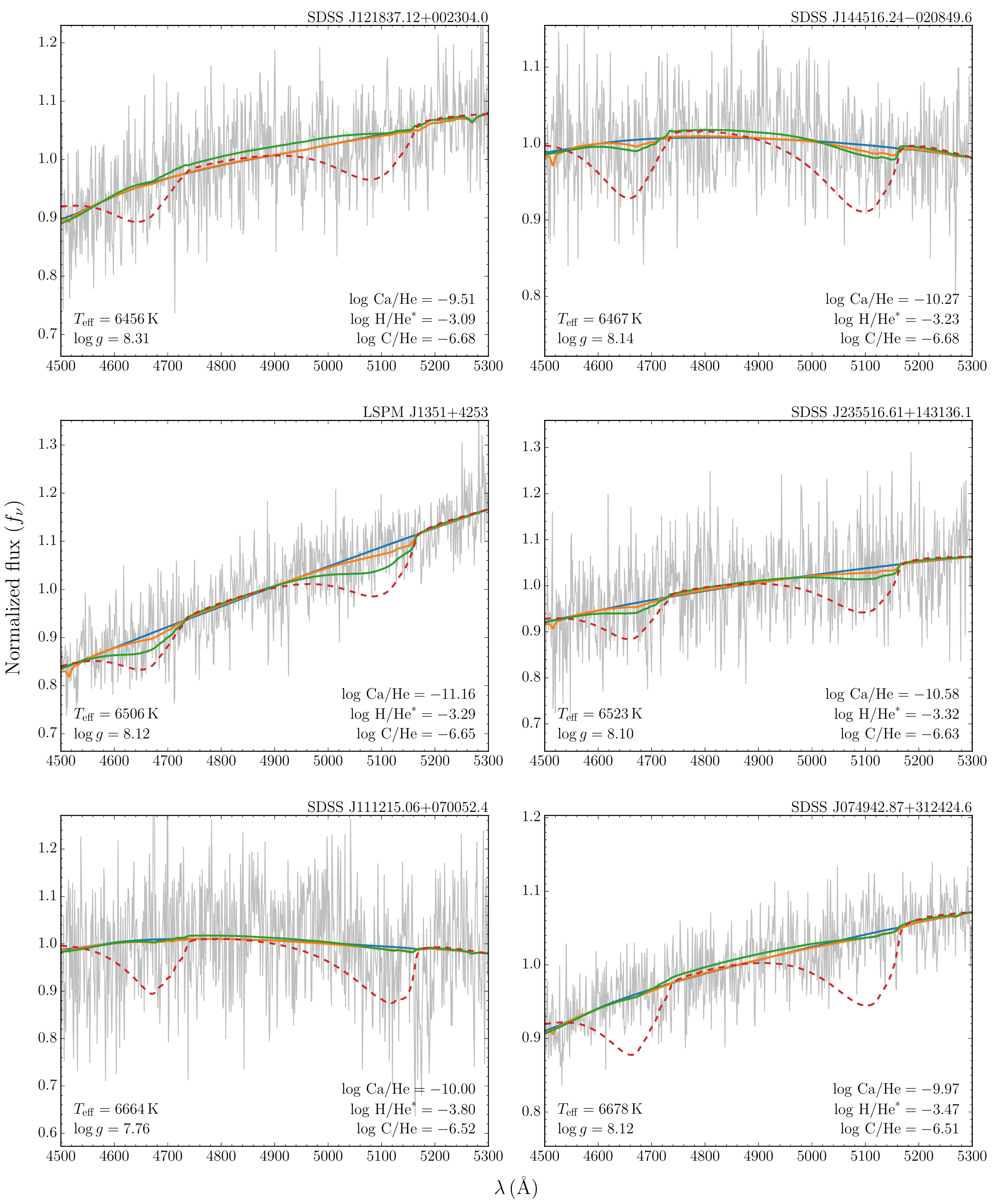}
    \caption{Comparison between SDSS spectra (grey) and different model atmospheres (described in the text, see Section~\ref{sec:implications}) convolved at the SDSS resolution. Panels for the 66 remaining objects are available in Figure~\ref{fig:panelapp}.}
    \label{fig:panel1}
\end{figure*}

To demonstrate that most DQs do transform directly into DZs, we have calculated model atmospheres for each of those 72~objects using the $T_{\rm eff}$, $\log g$, Ca/He and H/He values given in \cite{coutu2019}. In addition, we have computed models where we have boosted the carbon abundance so that C/He matches the expected carbon abundance for a DQ at that particular effective temperature (following the dashed red line of Figure~\ref{fig:DQscatter}). The resulting synthetic spectra are shown in Figure~\ref{fig:panel1}. For each object, we show four different models:
\begin{itemize}
    \item The blue model is a reference model computed assuming the same parameters as those found by \cite{coutu2019}. It has the same parameters as those given in each panel except for C/He, which is fixed so that C/Ca is chondritic.
    \item The orange model contains a trace of carbon typical of a DQ white dwarf at that temperature. It has the same parameters as those given in each panel.
    \item The green model is identical to the orange model, except that no trace of hydrogen was included in the calculations. In contrast, in \cite{coutu2019} and in the orange models, H/He is adjusted to reproduce the Balmer lines or to match the visibility limit of H$\alpha$ if no Balmer line is detected (in that case, an asterisk is displayed next to the hydrogen abundance in Figure~\ref{fig:panel1}).
    \item The red model (dashed line) is a DQ model with no polluting metals other than carbon. It has the same parameters as those given in each panel except that ${\rm H}/{\rm He}=0$ and ${\rm Ca}/{\rm He}=0$.
\end{itemize}
In all cases, a full model structure is recalculated: the synthetic spectra and underlying model atmospheres are fully consistent. The only parameters that were adjusted to the observations are the absolute scale and the slope of the synthetic spectra. All other parameters are fixed as described above. Finally, since \cite{coutu2019} adjusted Ca/He while keeping other abundance ratios constant (e.g., Mg/He), some objects with high Mg/Ca are not well represented by our models in the Mg$\,{\textsc{i}}$ 5168/5174/5185\,{\rm \AA} triplet region. In the context of this work, this is only a cosmetic issue that can safely be ignored.

The first thing to notice is that in most cases, the SDSS spectra have a sufficiently high signal-to-noise ratio (SNR) that Swan bands could be detected in DQ stars at those same $T_{\rm eff}$ and with typical carbon abundances (red dashed lines). In contrast, for the vast majority of objects, the same amount of carbon should not be detectable once we account for the presence of polluting metals (orange lines). \chg{For most stars, it would remain impossible to conclusively detect Swan bands even with observations with ${\rm SNR}=100$.} This conclusion still holds if we remove any hydrogen impurities from the model calculations (green lines). Hydrogen can further suppress the Swan bands by adding more free electrons to the atmosphere, which explains the difference between the orange and green models.

The key takeaway from Figure~\ref{fig:panel1} is that if, as expected, some objects in this sample are stars that have dredged up carbon from their deep interiors, they could not have been classified as DQZ because their Swan bands are too strongly suppressed to be detectable. The rarity of DQZ white dwarfs is not a surprise and is naturally explained by the well-known effect that electron donors have on the atmospheric structures of helium-dominated atmospheres. Figure~\ref{fig:panel1} also shows why it is unsurprising that the few known DQZ white dwarfs have very small accretion rates compared to the DZ population \citep{farihi2022}. The few objects where the Swan bands could have a chance of being detected with higher SNR observations are precisely those with the lowest levels of external pollution. \chg{For instance, we estimate that the Swan bands in the orange models of Figure~\ref{fig:panel1}/\ref{fig:panelapp} for the three most weakly polluted objects could be conclusively detected with an SNR $\gtrsim 40$ (WD~1005+030, $\log\,{\rm Ca/He}=-11.32$) and $\gtrsim 150$ (LSPM J1341+4253, $\log\,{\rm Ca/He}=-11.16$; USNO$-$B1.0 0937$-$00210798, $\log\,{\rm Ca/He}=-10.95$).}

So far, the hunt for more DQZs has been mostly focused on a search for metal lines (chiefly Ca$\,{\textsc{ii}}$ H \& K) in known DQ white dwarfs \citep{farihi2022}. By exclusively selecting known DQs, this survey strategy is effectively selecting objects that are very unlikely to be DQZs. In fact, since most known DQ white dwarfs have easily detectable Swan bands, they are objects that must have very low (if any) external pollution, otherwise they would not have strong Swan bands in the first place. A more promising survey strategy might be to look for very shallow Swan bands in known DZ white dwarfs with very low Ca/He, since a low external pollution implies a weaker Swan bands suppression.

\section{Conclusion}
\label{sec:conclusion}
Using state-of-the-art model atmospheres, we have shown that a classical DQ star with a typical carbon abundance directly transforms into a DZ white dwarf following the accretion of a typical amount of metals from planetary debris. The accreted material decreases the density of the atmosphere, which results in a smaller C$_2$ abundance and a strong suppression of the Swan bands. This naturally explains the observed paucity of DQZ stars as well as the very small accretion rates inferred for known DQZs.

Those findings nullify the main argument put forward by \cite{farihi2022} to support the idea that all DQ stars are the product of binary evolution. We cannot definitely rule out this scenario, but our conclusions at least lessen the need for this hypothesis. That being said, we recognize that there are still other properties of DQ stars that remain hard to explain (notably the observed deficit of unevolved companions in post-common envelope binaries) and each of those should be further scrutinized.

\begin{acknowledgements}
SB thanks Pierre Bergeron, Patrick Dufour and Antoine B\'edard for useful discussions. SB is a Banting Postdoctoral Fellow and a CITA National Fellow, supported by the Natural Sciences and Engineering Research Council of Canada (NSERC). This work has made use of the Montreal White Dwarf Database \citep{dufour2017}.
\end{acknowledgements}

\bibliographystyle{aa}
\bibliography{references}

\begin{appendix}
\section{Supplementary figures}
\FloatBarrier

\begin{figure}[H]
	\includegraphics[width=\columnwidth]{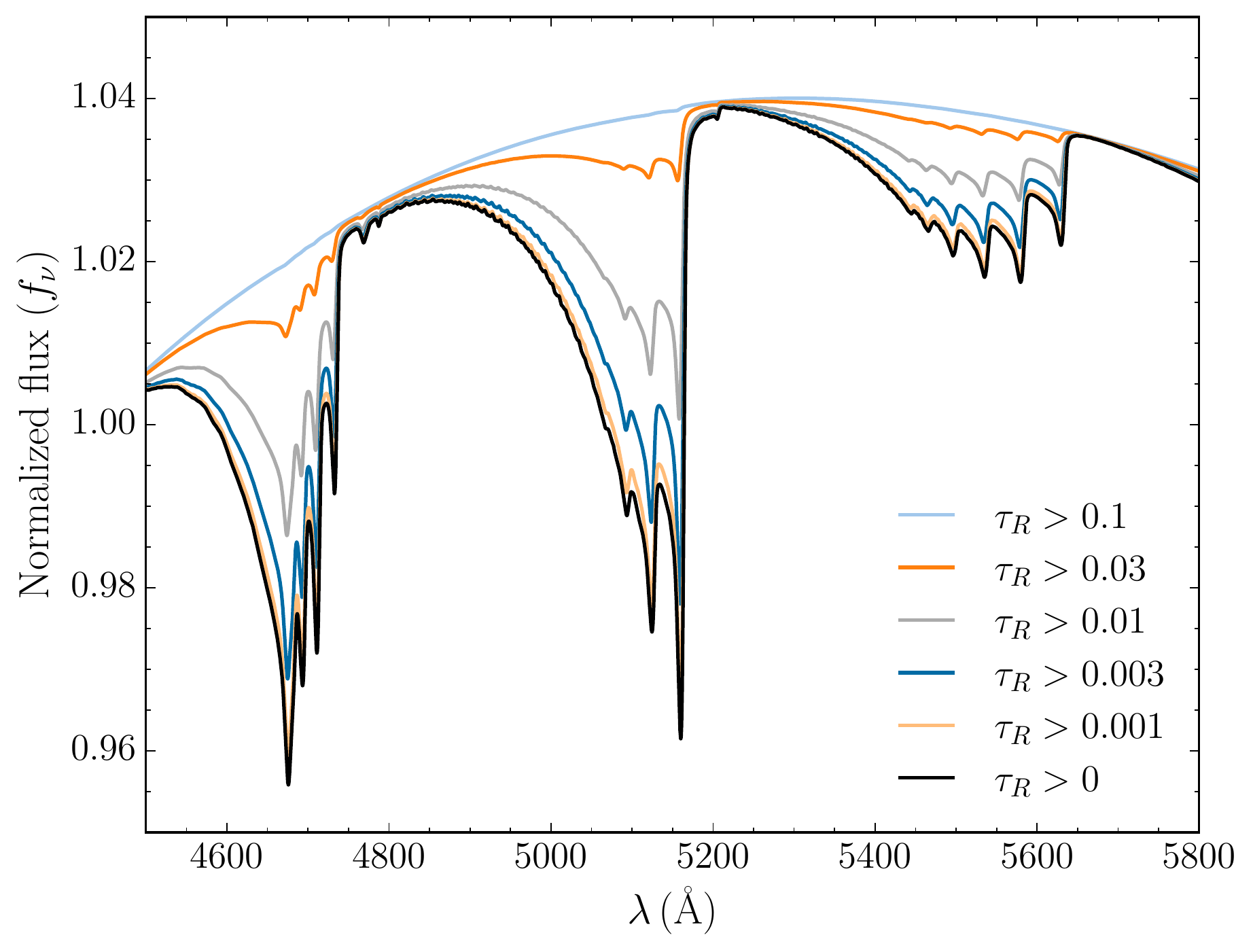}
    \caption{Synthetic spectra of a DQZ white dwarf with $T_{\rm eff}=8000\,$K, $\log g=8$, $\log\,{\rm C/He}=-5.5$ and $\log\,{\rm Ca/He}=-11.5$. Each spectrum was computed using the same unperturbed model structure, but the Swan bands opacity was omitted from the spectrum calculation for Rosseland optical depths smaller than the value given in the legend. For example, for the grey spectrum, the Swan bands opacity was only included for atmospheric layers deeper than $\tau_R=0.01$.}
    \label{fig:taucontrib}
\end{figure}

\begin{figure}[H]
	\includegraphics[width=\columnwidth]{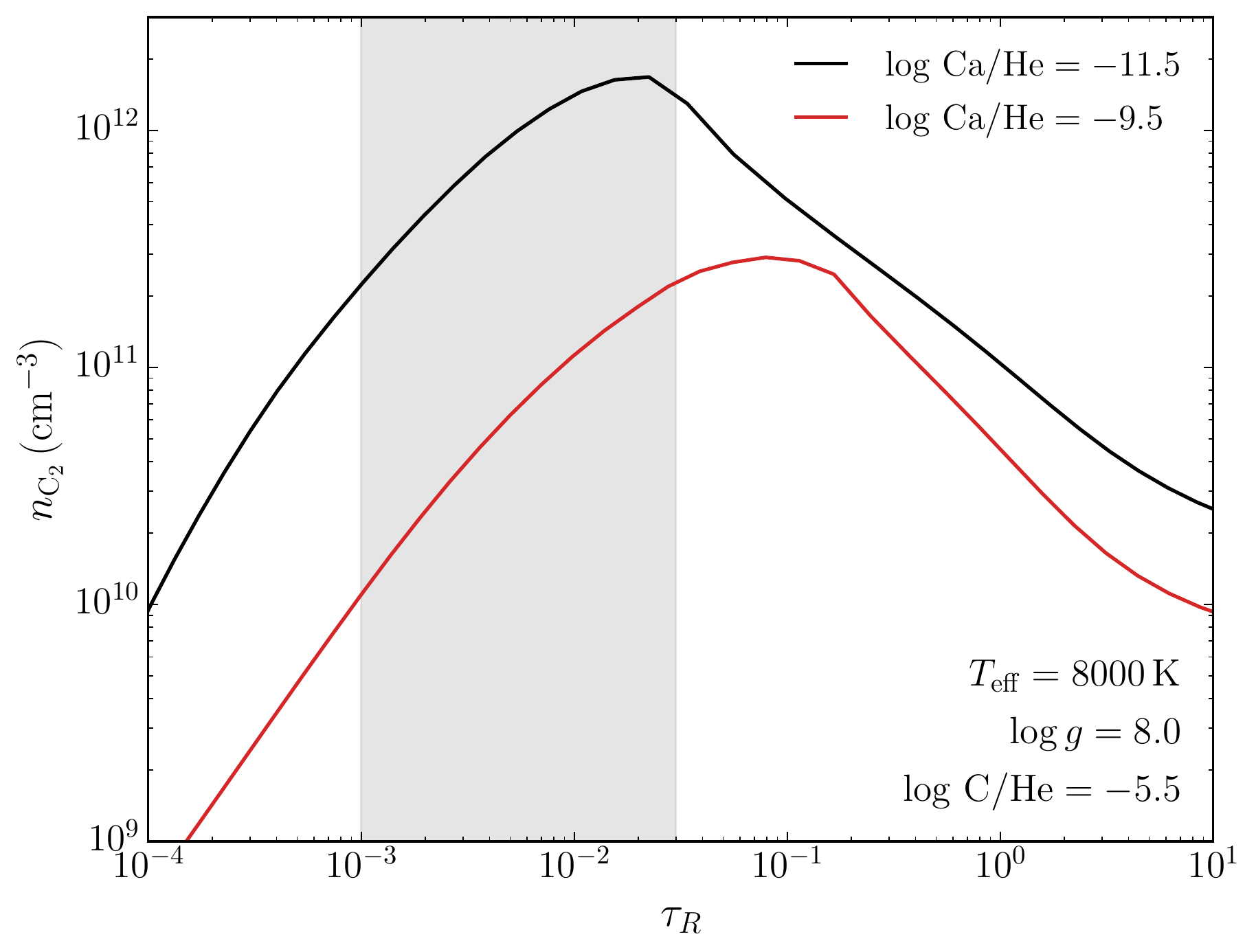}
    \caption{Molecular carbon number density as a function of Rosseland mean optical depth for DQZ white dwarfs with $T_{\rm eff}=8000\,$K, $\log g=8$, $\log\,{\rm C/He}=-5.5$ and different Ca/He values (the black line corresponds to $\log\,{\rm Ca/He}=-11.5$ and the red line to $\log\,{\rm Ca/He}=-9.5$).}
    \label{fig:C2abn}
\end{figure}

\FloatBarrier

\begin{figure}
	\includegraphics[width=1\columnwidth]{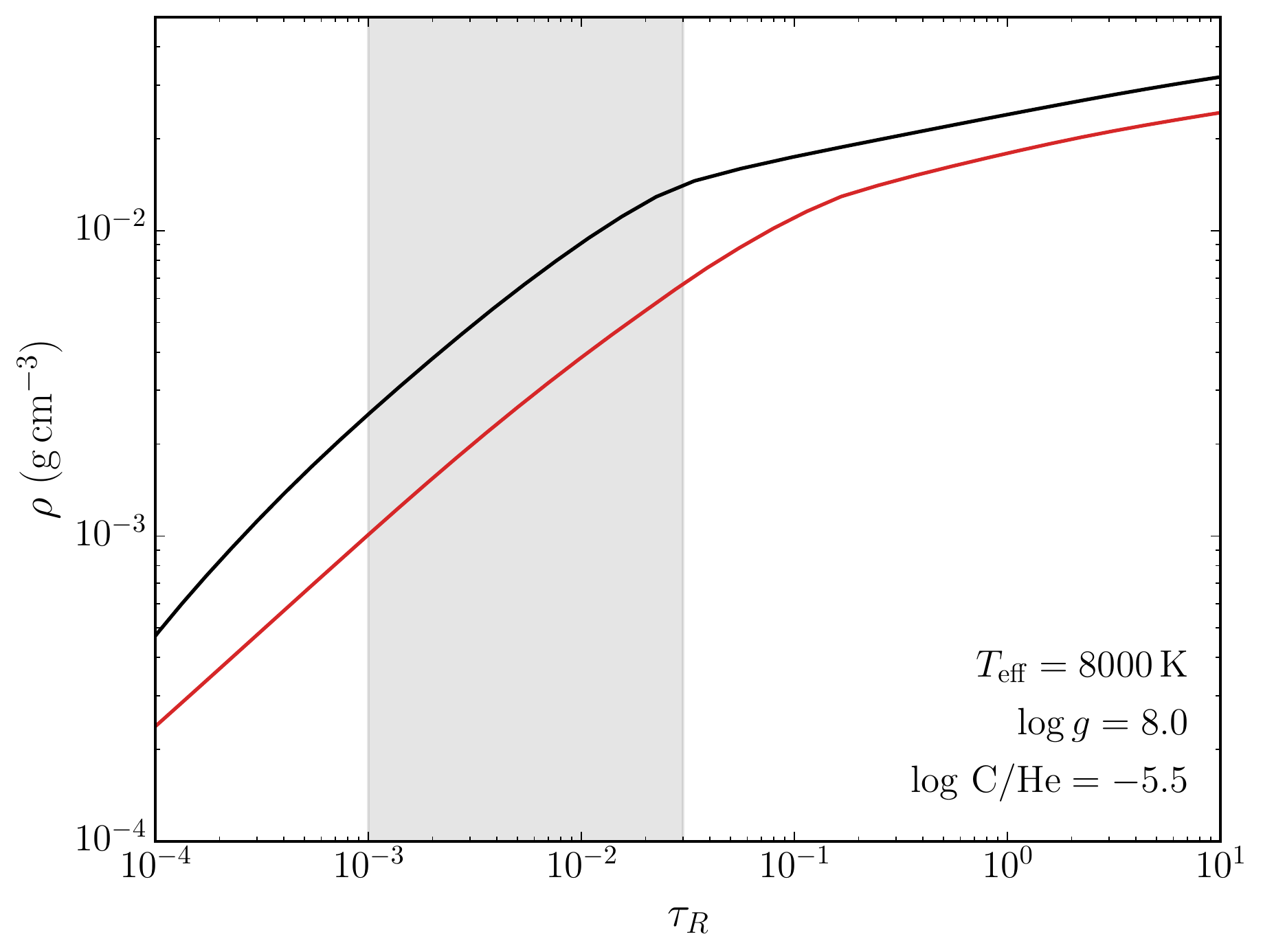}
	\includegraphics[width=1\columnwidth]{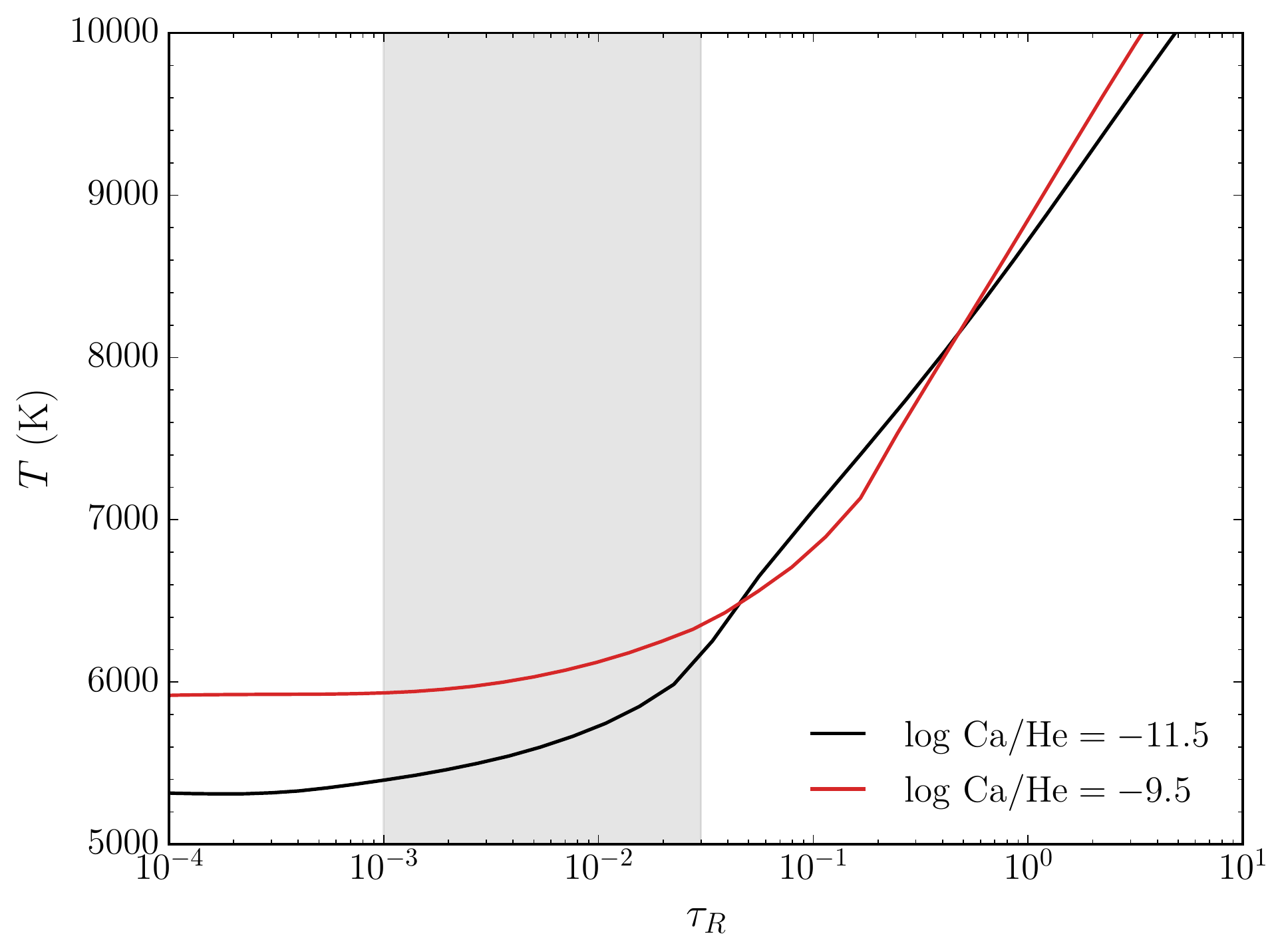}
    \caption{Mass density (top panel) and temperature (bottom panel) as a function of Rosseland mean optical depth for DQZ white dwarfs with $T_{\rm eff}=8000\,$K, $\log g=8$, $\log\,{\rm C/He}=-5.5$ and different Ca/He values (black lines correspond to $\log\,{\rm Ca/He}=-11.5$ and red lines to $\log\,{\rm Ca/He}=-9.5$).}
    \label{fig:structure}
\end{figure}

\begin{figure}
	\includegraphics[width=1\columnwidth]{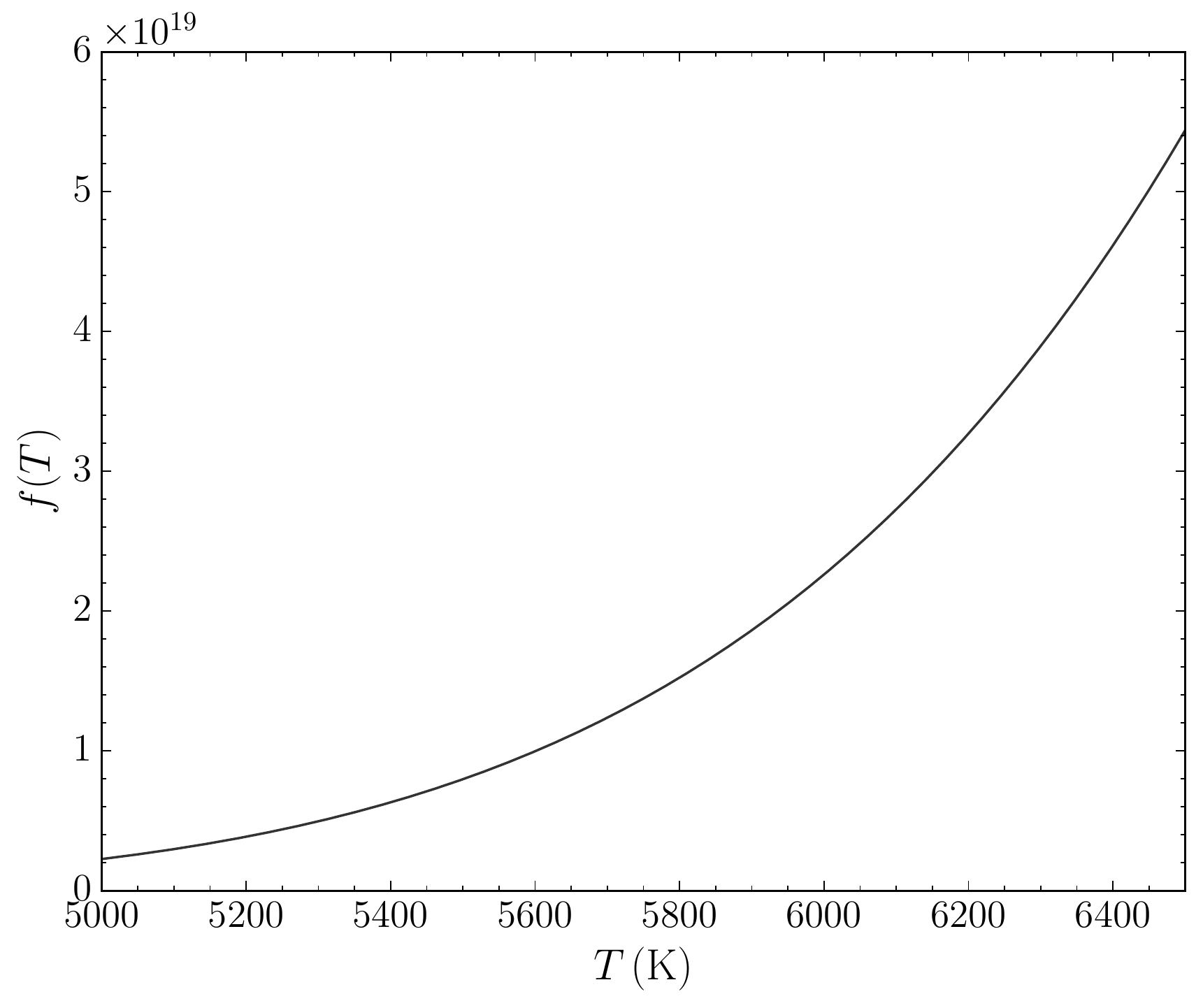}
    \caption{Right-hand side of Equation~\eqref{eq:C2}. The partition functions given in \protect\cite{irwin1981} were used for this exercise.}
    \label{fig:fT}
\end{figure}

\FloatBarrier

\begin{figure*}
    \centering
	\includegraphics[width=2\columnwidth]{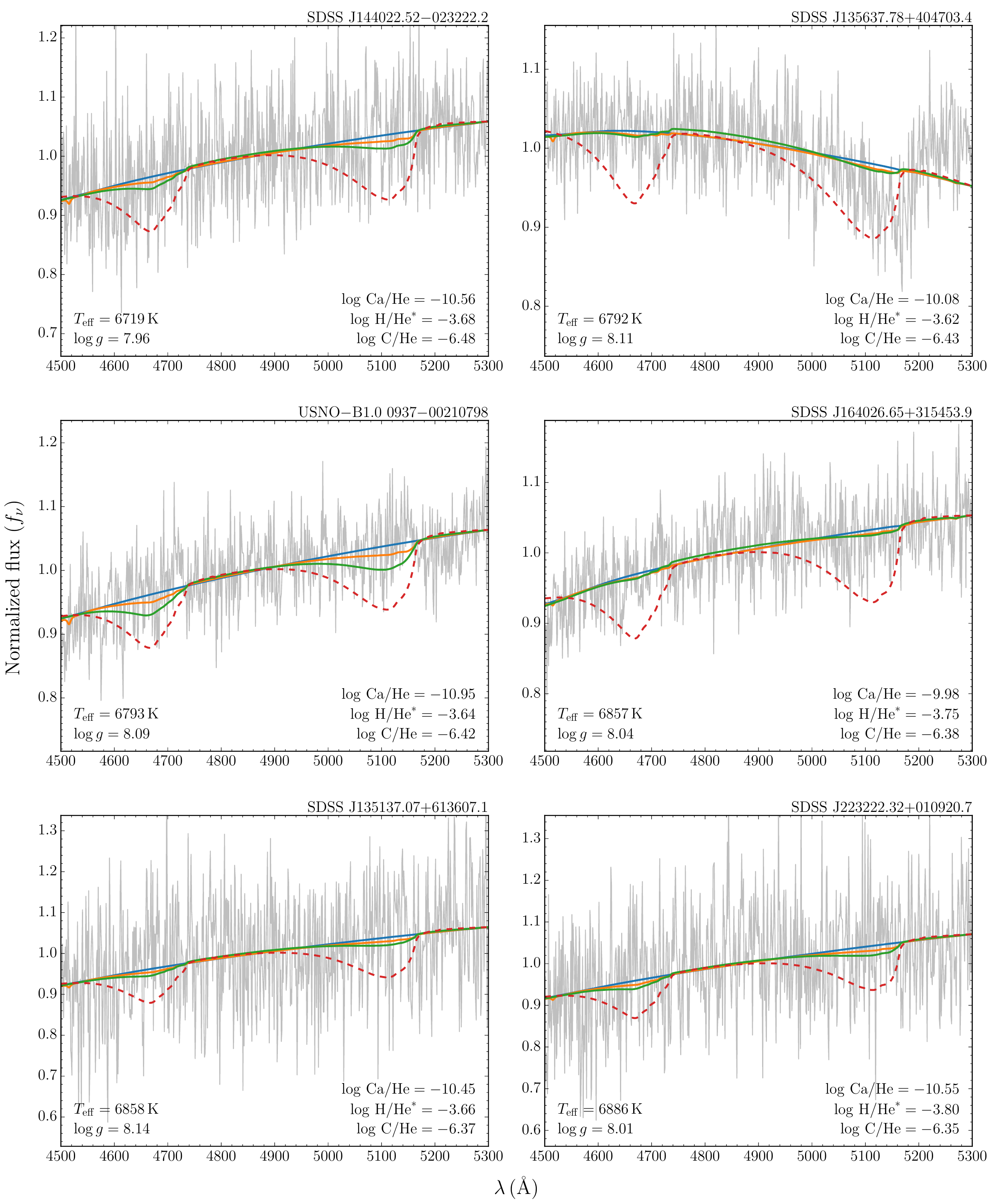}
    \caption{Comparison between SDSS spectra (grey) and different model atmospheres (described in the text, see Section~\ref{sec:implications}) convolved at the SDSS resolution. Continued from Figure~\ref{fig:panel1}.}
    \label{fig:panelapp}
\end{figure*}

\addtocounter{figure}{-1}
\begin{figure*}
    \centering
	\includegraphics[width=2\columnwidth]{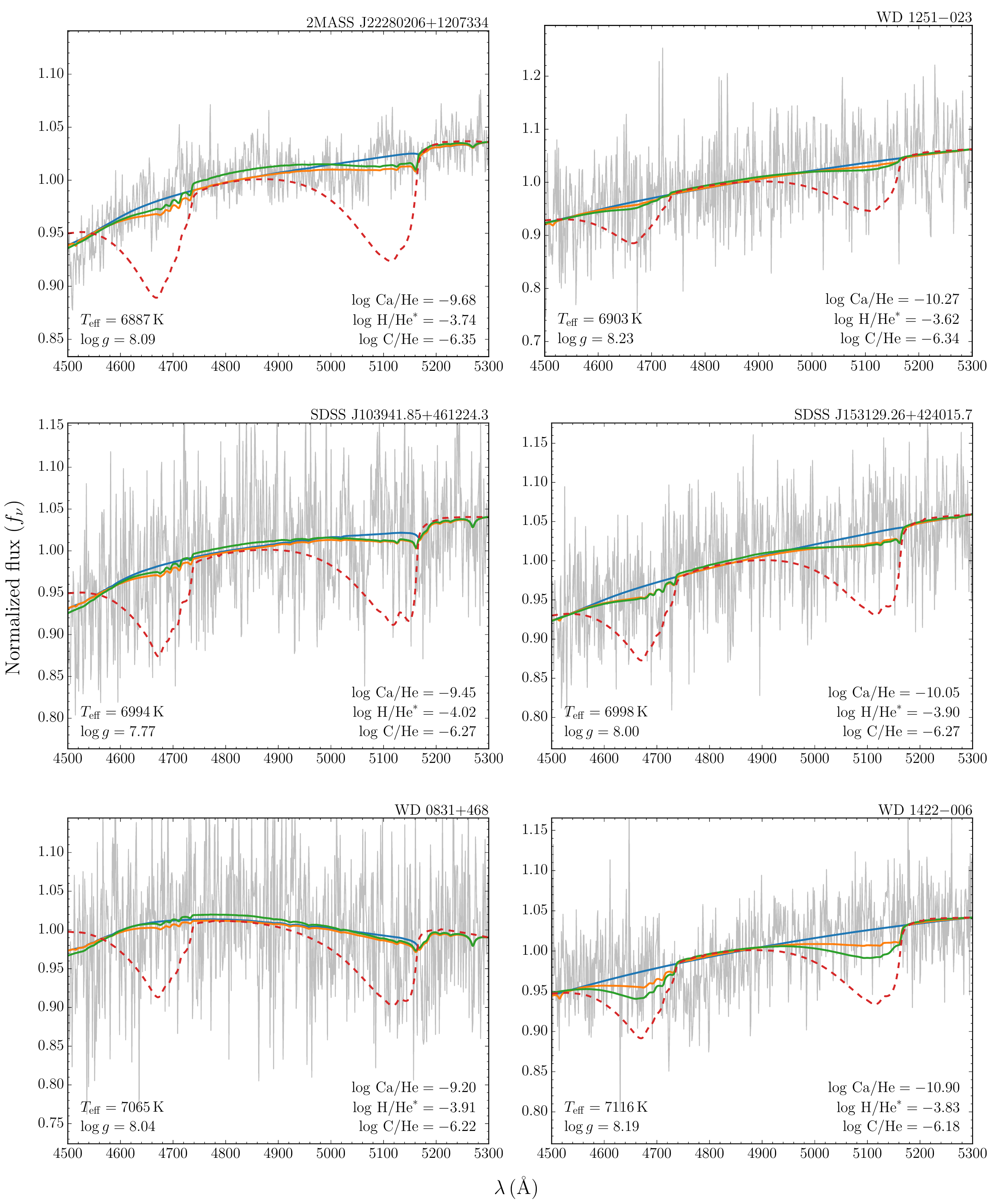}
    \caption{continued.}
\end{figure*}

\addtocounter{figure}{-1}
\begin{figure*}
    \centering
	\includegraphics[width=2\columnwidth]{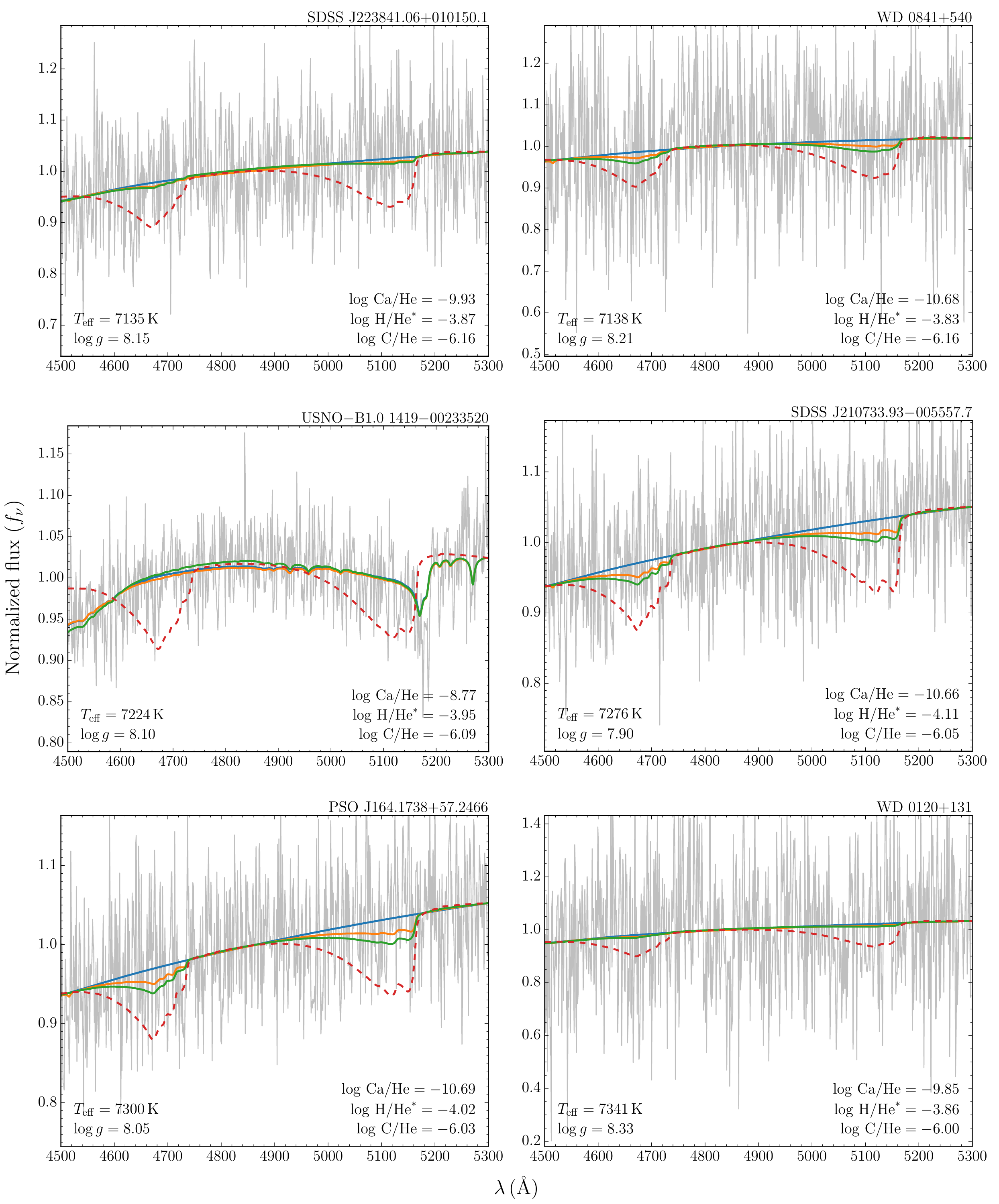}
    \caption{continued.}
\end{figure*}

\addtocounter{figure}{-1}
\begin{figure*}
    \centering
	\includegraphics[width=2\columnwidth]{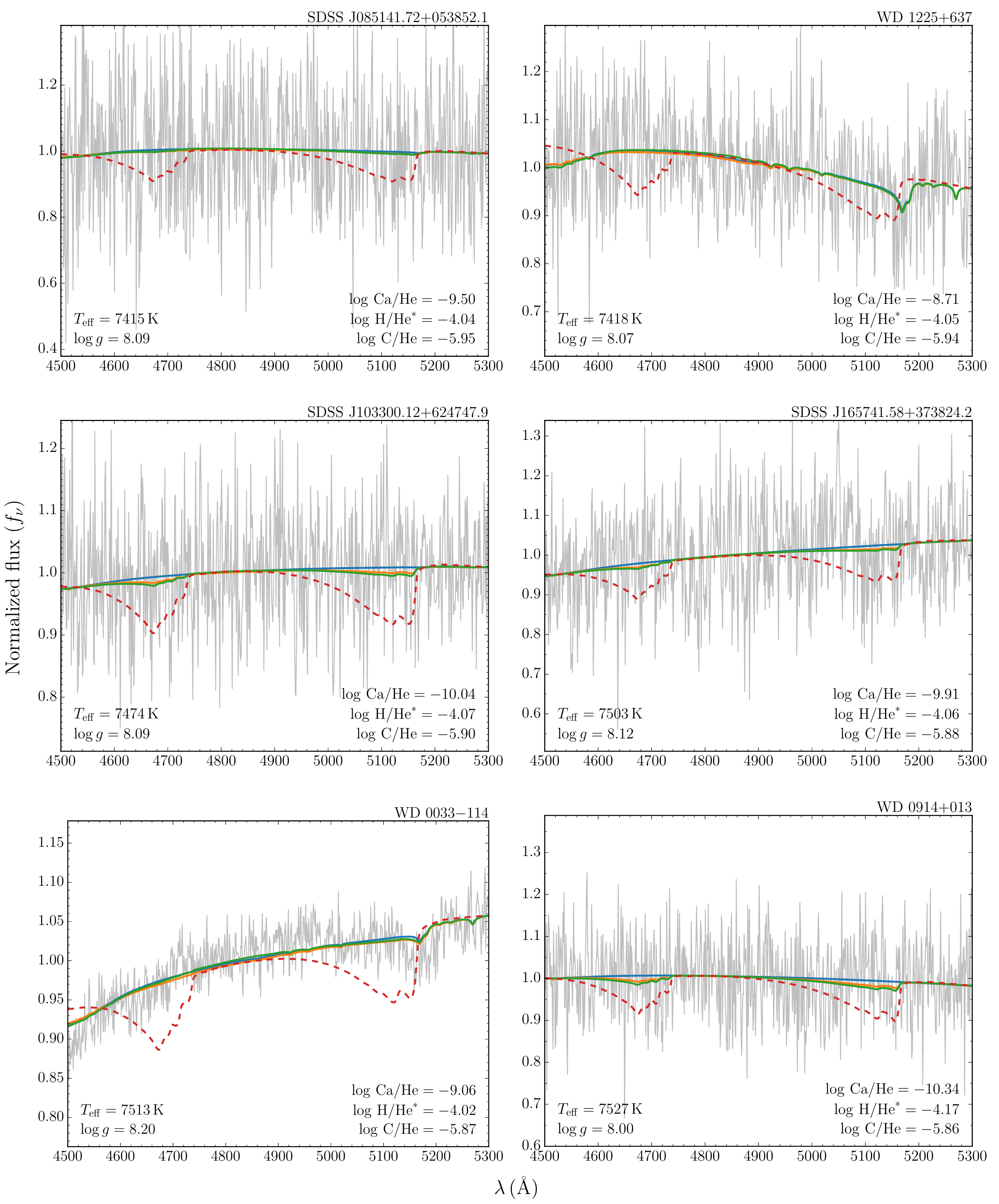}
    \caption{continued.}
\end{figure*}

\addtocounter{figure}{-1}
\begin{figure*}
    \centering
	\includegraphics[width=2\columnwidth]{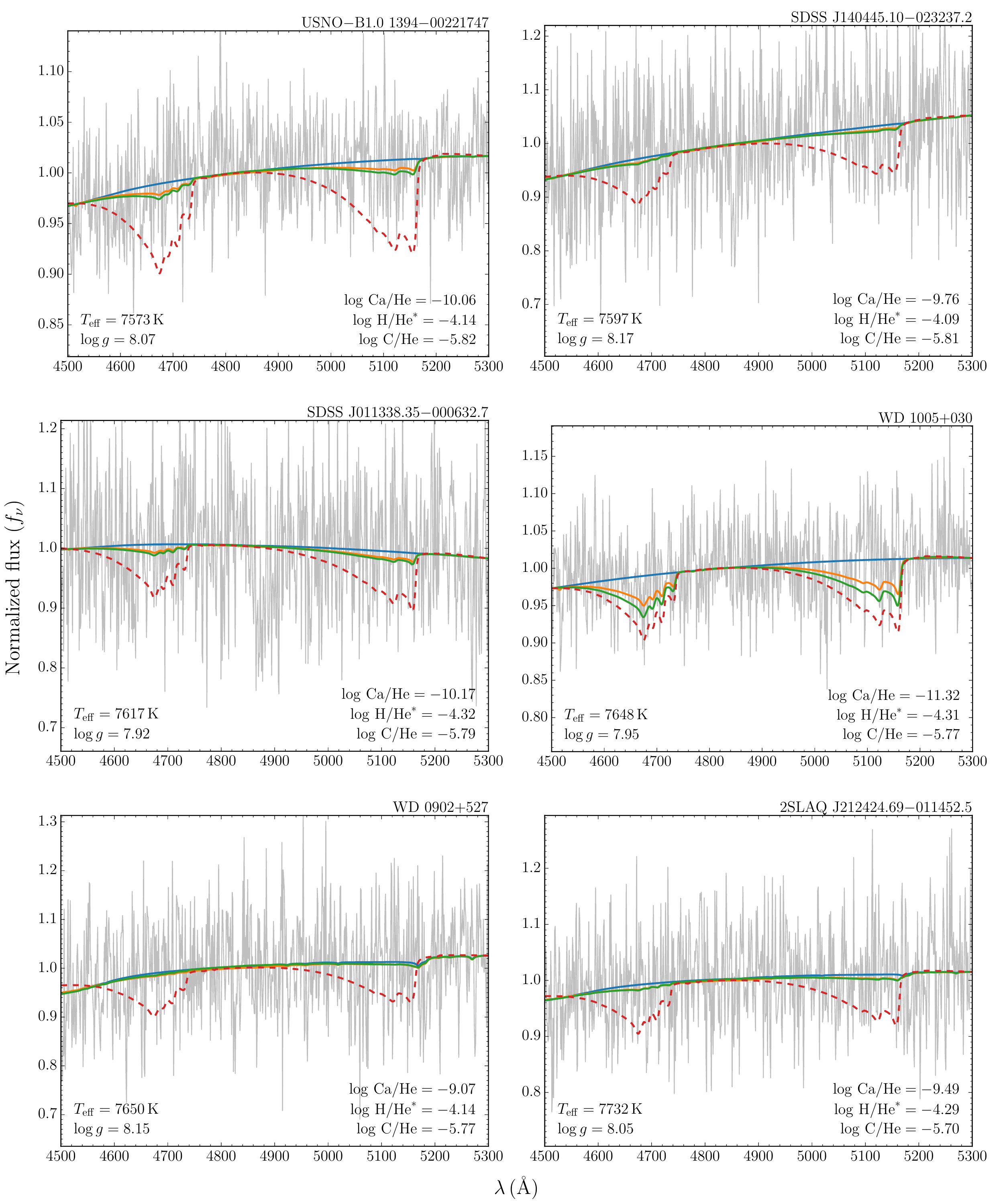}
    \caption{continued.}
\end{figure*}

\addtocounter{figure}{-1}
\begin{figure*}
    \centering
	\includegraphics[width=2\columnwidth]{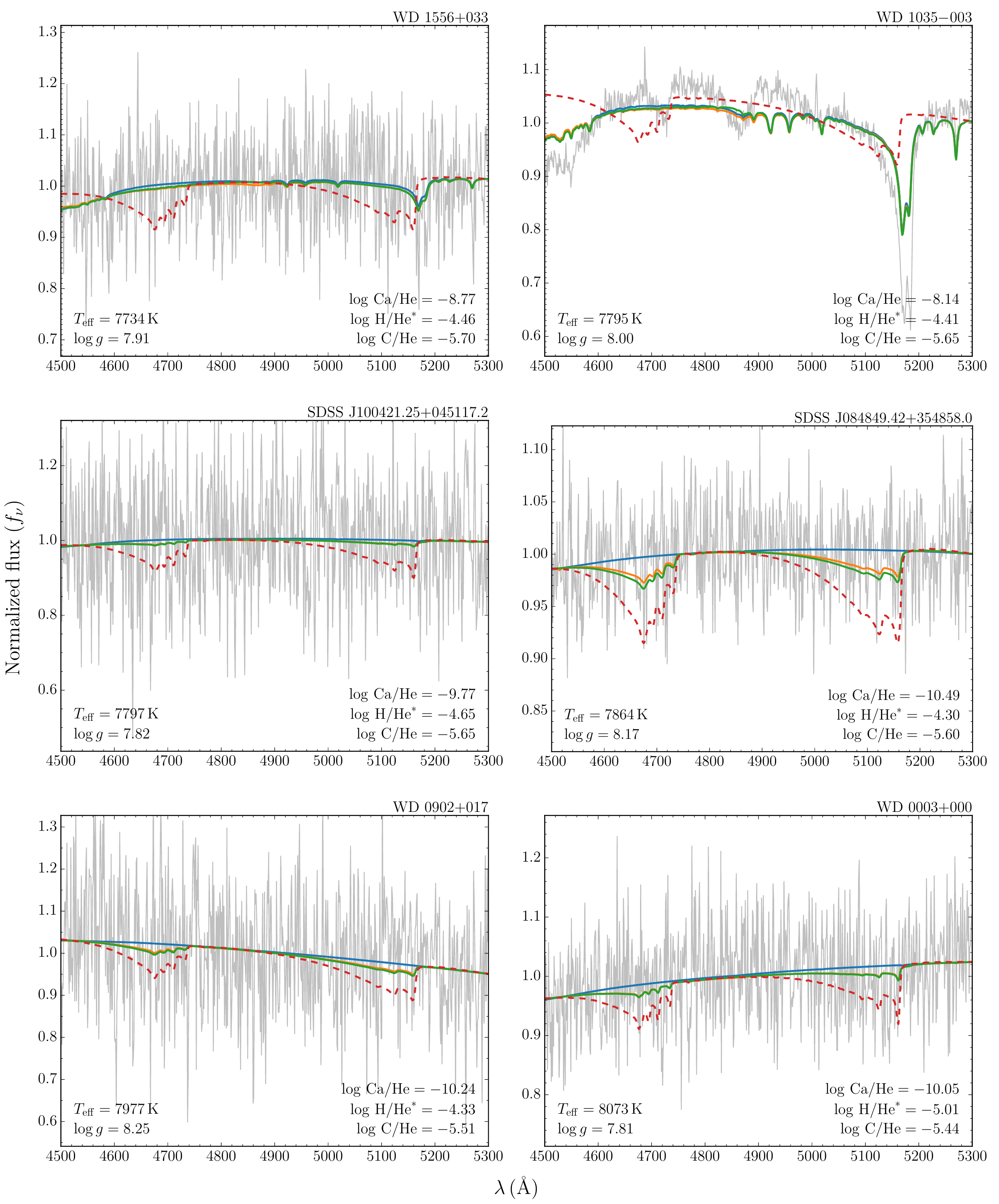}
    \caption{continued.}
\end{figure*}

\addtocounter{figure}{-1}
\begin{figure*}
    \centering
	\includegraphics[width=2\columnwidth]{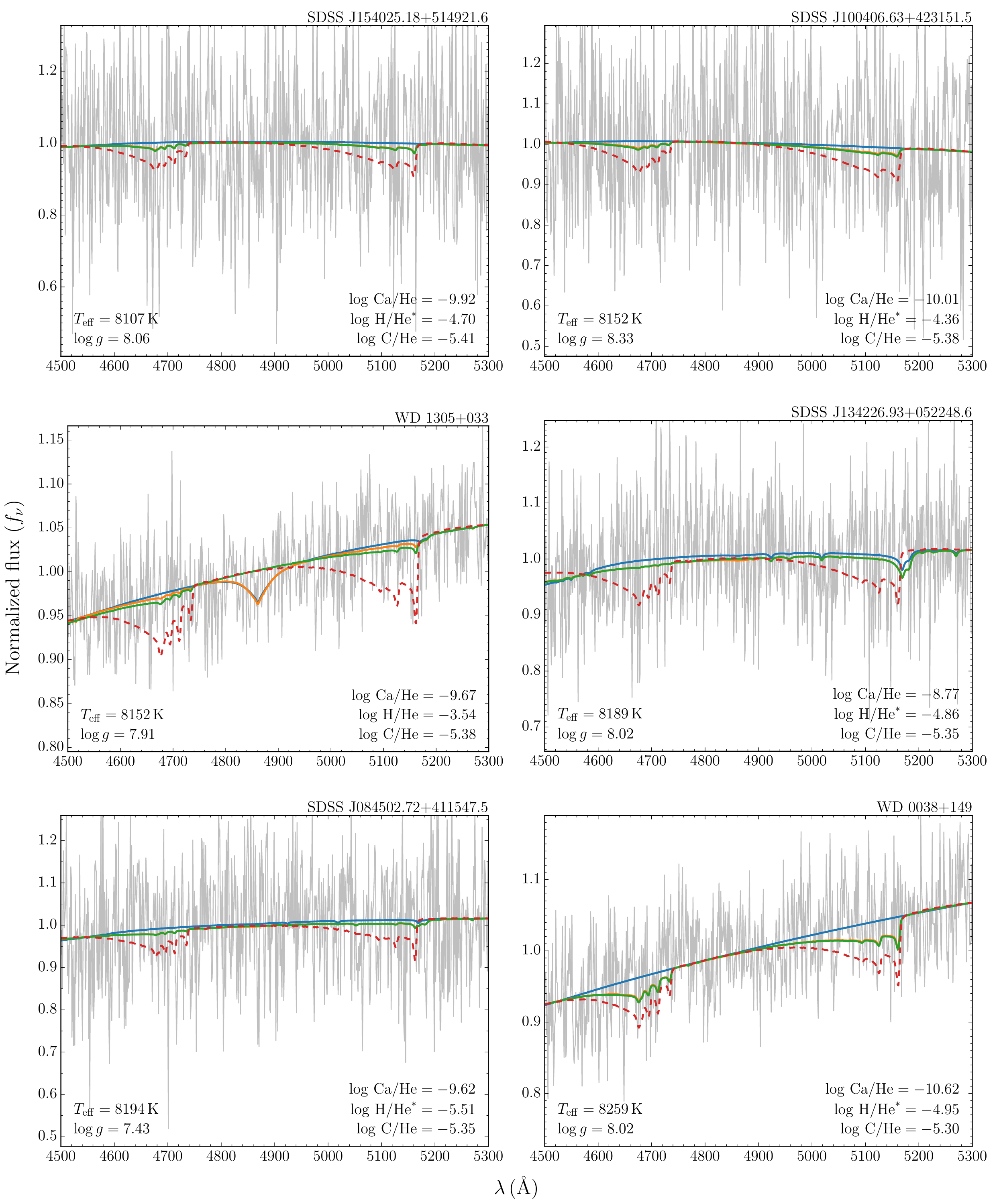}
    \caption{continued.}
\end{figure*}

\addtocounter{figure}{-1}
\begin{figure*}
    \centering
	\includegraphics[width=2\columnwidth]{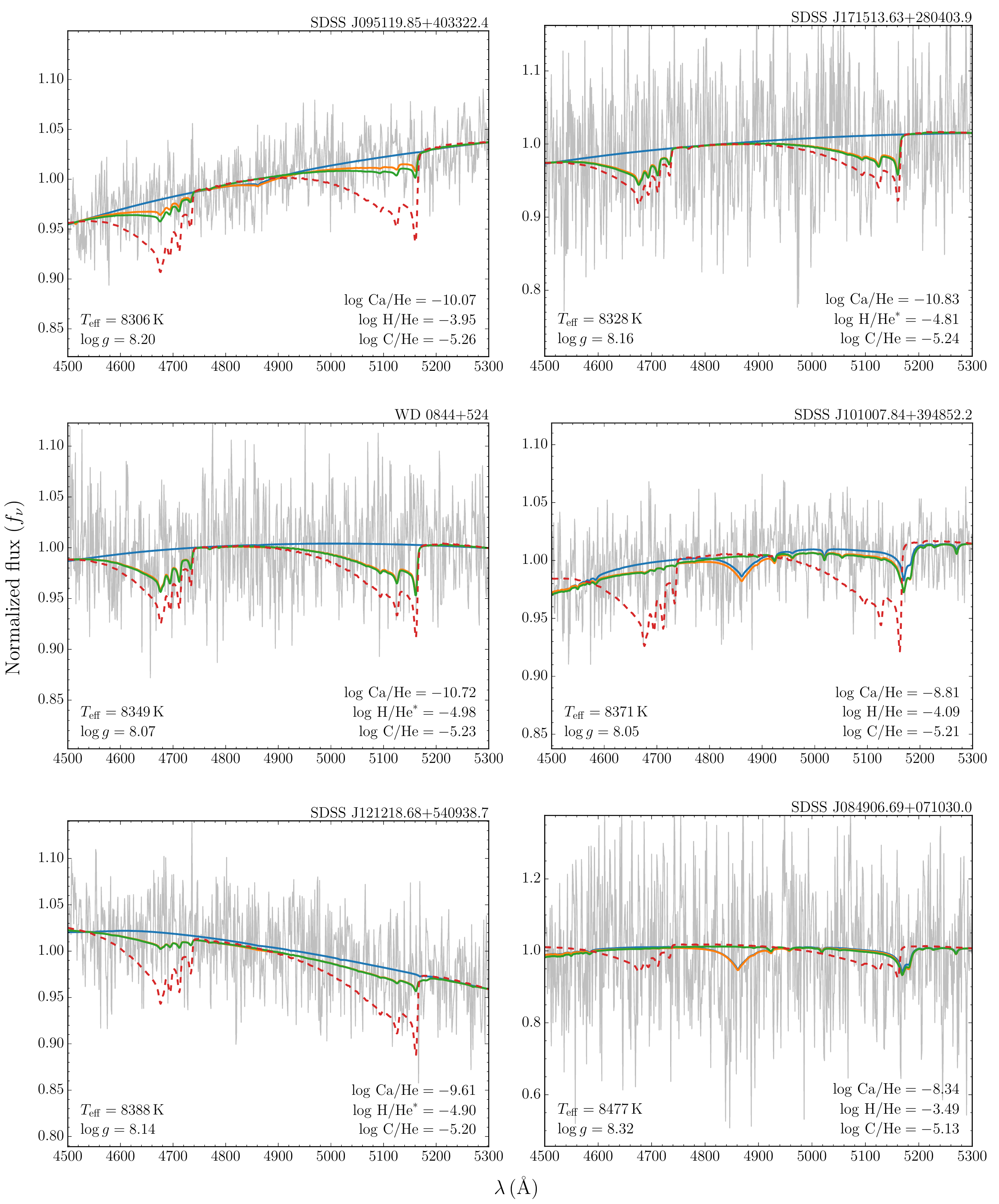}
    \caption{continued.}
\end{figure*}

\addtocounter{figure}{-1}
\begin{figure*}
    \centering
	\includegraphics[width=2\columnwidth]{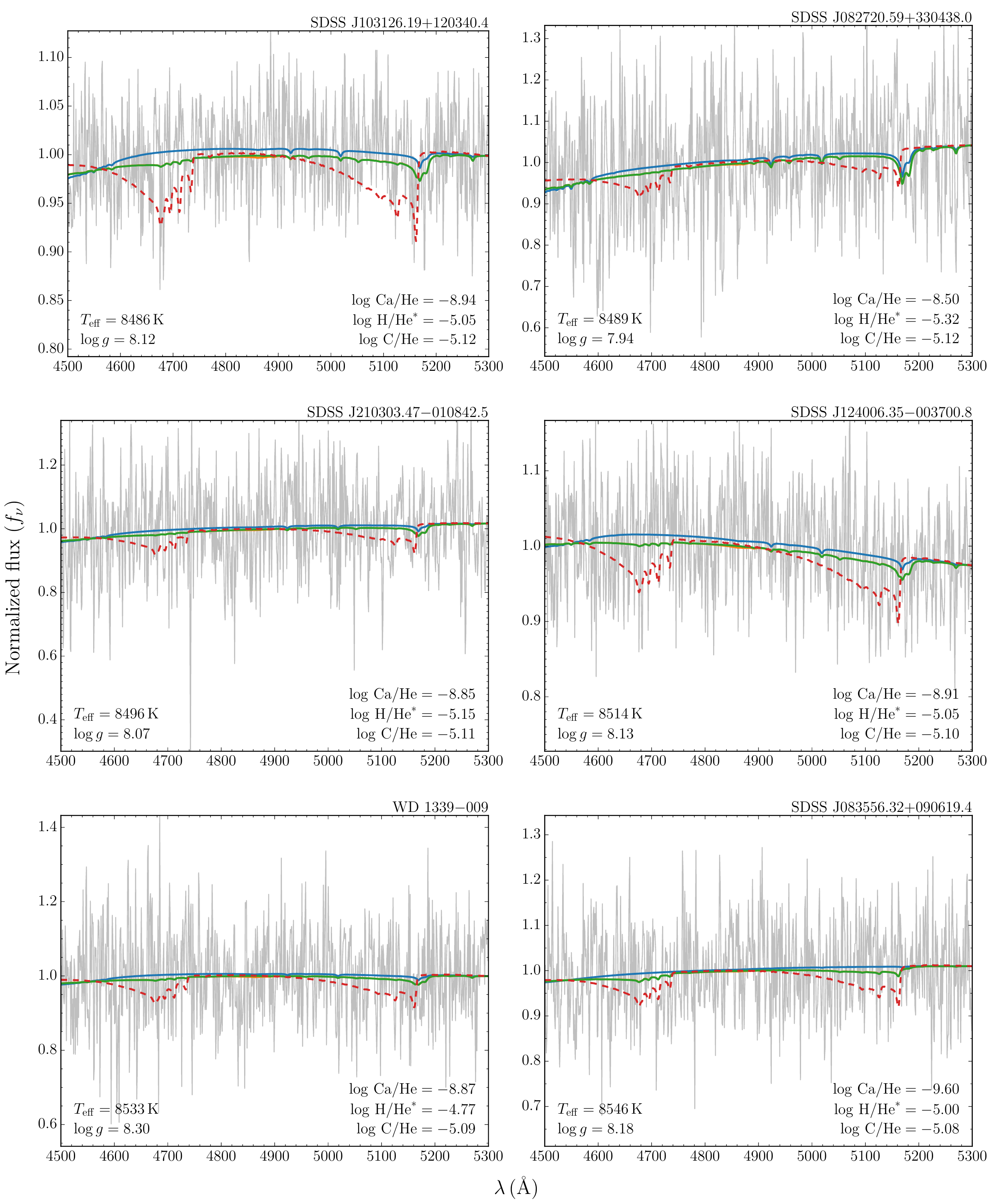}
    \caption{continued.}
\end{figure*}

\addtocounter{figure}{-1}
\begin{figure*}
    \centering
	\includegraphics[width=2\columnwidth]{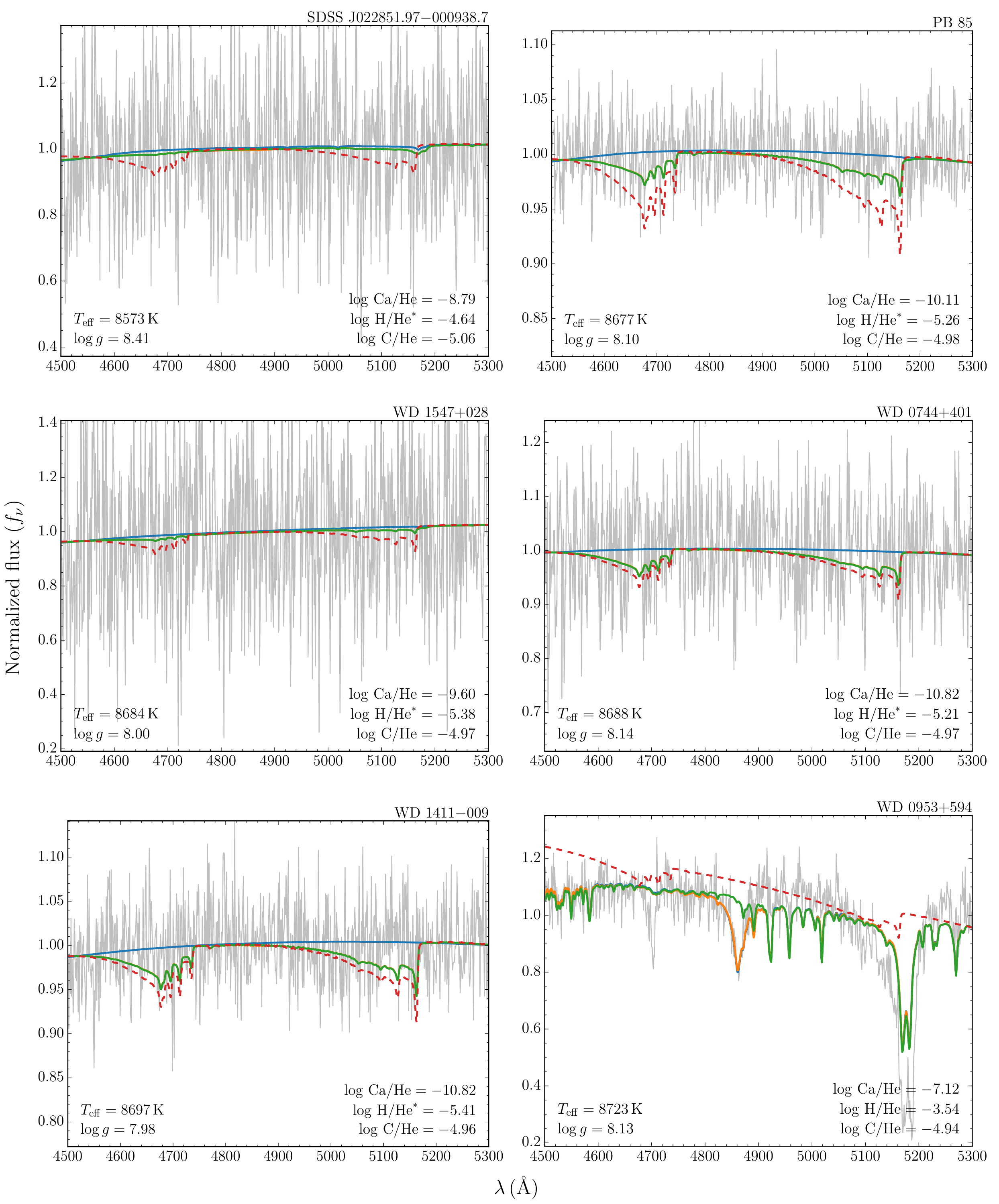}
    \caption{continued.}
\end{figure*}

\addtocounter{figure}{-1}
\begin{figure*}
    \centering
	\includegraphics[width=2\columnwidth]{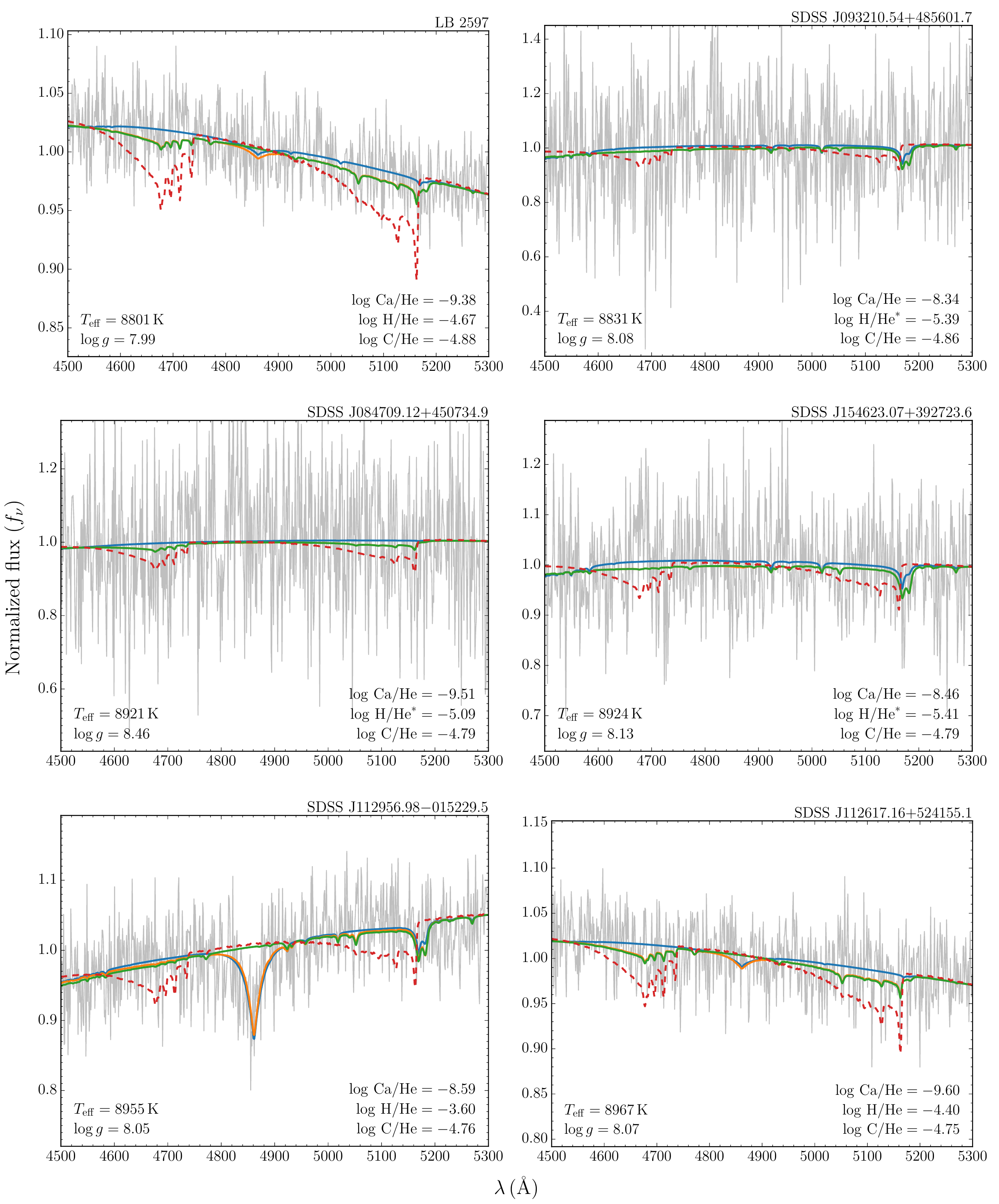}
    \caption{continued.}
\end{figure*}

\end{appendix}

\end{document}